
\input amssym.def
\input amssym.tex

\magnification=1200
\baselineskip=14pt plus 1pt minus 1pt
\tolerance=1000\hfuzz=1pt


\font\bigfont=cmr10 scaled\magstep3

\def\section#1#2{\vskip32pt plus4pt \goodbreak \noindent{\bf#1. #2}
        \xdef\currentsec{#1} \global\eqnum=0 \global\thmnum=0}

\def\leftheader{}
\def\rightheader{}
\def\header{\headline={\tenrm\ifnum\pageno>1
                        \ifodd\pageno\rightheader
                         \else\noindent\leftheader
                          \fi\else\hfil\fi}}

\newcount\thmnum
\global\thmnum=0
\def\prop#1#2{\global\advance\thmnum by 1
        \xdef#1{Proposition \currentsec.\the\thmnum}
        \bigbreak\noindent{\bf Proposition \currentsec.\the\thmnum.}
        {\it#2} }
\def\define#1#2{\global\advance\thmnum by 1
        \xdef#1{Definition \currentsec.\the\thmnum}
        \bigbreak\noindent{\bf Definition \currentsec.\the\thmnum.}
        {\it#2} }
\def\lemma#1#2{\global\advance\thmnum by 1
        \xdef#1{Lemma \currentsec.\the\thmnum}
        \bigbreak\noindent{\bf Lemma \currentsec.\the\thmnum.}
        {\it#2}}
\def\thm#1#2{\global\advance\thmnum by 1
        \xdef#1{Theorem \currentsec.\the\thmnum}
        \bigbreak\noindent{\bf Theorem \currentsec.\the\thmnum.}
        {\it#2} }
\def\cor#1#2{\global\advance\thmnum by 1
        \xdef#1{Corollary \currentsec.\the\thmnum}
        \bigbreak\noindent{\bf Corollary \currentsec.\the\thmnum.}
        {\it#2} }
\def\conj#1#2{\global\advance\thmnum by 1
        \xdef#1{Conjecture \currentsec.\the\thmnum}
        \bigbreak\noindent{\bf Conjecture \currentsec.\the\thmnum.}
        {\it#2} }

\newcount\eqnum
\global\eqnum=0
\def\num{\global\advance\eqnum by 1
        \eqno({\rm\currentsec}.\the\eqnum)}
\def\eqalignnum{\global\advance\eqnum by 1
        ({\rm\currentsec}.\the\eqnum)}
\def\ref#1{\num  \xdef#1{(\currentsec.\the\eqnum)}}
\def\eqalignref#1{\eqalignnum  \xdef#1{(\currentsec.\the\eqnum)}}

\def\title#1{\centerline{\bf\bigfont#1}}

\newcount\subnum
\def\Alph#1{\ifcase#1\or A\or B\or C\or D\or E\or F\or G\or H\fi}
\def\subsec{\global\advance\subnum by 1
        \vskip12pt plus4pt \goodbreak \noindent
        {\bf \currentsec.\Alph\subnum.}  }
\def\newsubsec{\global\subnum=1 \vskip6pt\noindent
        {\bf \currentsec.\Alph\subnum.}  }
\def\today{\ifcase\month\or January\or February\or March\or
        April\or May\or June\or July\or August\or September\or
        October\or November\or December\fi\space\number\day,
        \number\year}

\def\ol{\overline}

\def\tr{\mathop{\rm tr}\nolimits}

\def\bC{{\Bbb C}}
\def\bN{{\Bbb N}}

\def\bR{{\Bbb R}}
\def\bT{{\Bbb T}}
\def\bZ{{\Bbb Z}}
\def\cA{{\cal A}}
\def\cB{{\cal B}}

\def\cH{{\cal H}}

\def\cM{{\cal M}}
\def\cO{{\cal O}}

\def\cS{{\cal S}}

\def\fA{{\frak A}}

\def\fN{{\frak N}}
\def\fP{{\frak P}}
\def\fR{{\frak R}}

\def\fZ{{\frak Z}}
\chardef\o="1C
\def\frac#1#2{{#1\over #2}}

\header
\def\leftheader{\centerline{S. KLIMEK and A. LESNIEWSKI}}
\def\rightheader{\centerline{QUANTIZED CHAOTIC DYNAMICS}}

\def\intsec{I}
\def\planesec{II}
\def\torussec{III}
\def\bakersec{IV}
\def\connessec{V}
\def\entropysec{VI}

\def\hil{{\cal H}^2(\bC,d\mu_\hbar)}
\def\bak{B_\hbar}

{\baselineskip=12pt
\nopagenumbers
\line{\hfill \bf HUTMP 95/441}
\line{\hfill October 10, 1995}
\line{\hfill to appear in {\it Annals of Physics}}
\vfill
\title{Quantized chaotic dynamics}
\vskip 1cm
\title{and non-commutative KS entropy}
\vskip 1in
\centerline{{\bf S\l awomir Klimek}$^*$
\footnote{$^1$}{Supported in part by the National Science Foundation under
grants DMS--9206936 and DMS--9500463}
and {\bf Andrzej Le\'sniewski}$^{**}$\footnote{$^2$}
{Supported in part by the Department of Energy under grant
DE--FG02--88ER25065 and by the National Science Foundation under grant
DMS--9424344}}
\vskip 12pt
\centerline{$^*$Department of Mathematics}
\centerline{IUPUI}
\centerline{Indianapolis, IN 46205, USA}
\vskip 12pt
\centerline{$^{**}$Lyman Laboratory of Physics}
\centerline{Harvard University}
\centerline{Cambridge, MA 02138, USA}
\vskip 1in\noindent
{\bf Abstract}. We study the quantization of two examples of
classically chaotic dynamics, the Anosov dynamics of ``cat maps''
on a two dimensional torus, and the dynamics of baker's maps.
Each of these dynamics is implemented as a discrete group of
automorphisms of a von Neumann algebra of functions on a quantized
torus. We compute the non-commutative generalization of the
Kolmogorov-Sinai entropy, namely the Connes-St\o rmer entropy,
of the generator of this group, and find that its value is equal
to the classical value. This can be interpreted as a sign of
persistence of chaotic behavior in a dynamical system under
quantization.
\vfill\eject}

\section\intsec{Introduction}

\newsubsec
One of the characteristic features of chaos in classical dynamics
is the positivity of the Kolmogorov-Sinai (KS) entropy.	The KS
entropy is a natural measure of mixing in phase space resulting
from the time evolution of a dynamical system. Indeed, one can adopt
the positivity of the KS entropy as a convenient way of defining
chaos in a classical dynamical system. Through Pesin's theorem,
this is related to another characteristic feature of chaotic
evolution, namely the positivity of Lyapunov exponents.

The focus of the emerging field of ``quantum chaology'' [B2], [HT],
[N], [V2], is the study of quantum dynamics arising from quantization
of classically chaotic systems.	Much emphasis has been put on understanding
the semiclassical approximation to the actual quantum dynamics, and
it is, in fact, a somewhat controversial issue whether ``quantum
chaos'' exists beyond this approximation.

In this paper we propose that a natural quantity to exhibit
quantum chaos in a class of quantized dynamics is the positivity
of the Connes-St\o rmer (CS) entropy. The CS entropy is defined in
the context of von Neumann algebras, and is a natural extension of
the KS entropy to the non-commutative context. We focus our attention
on examples of quantized dynamics on a torus, namely the dynamics of
quantized cat maps and the dynamics of quantized baker's maps, and show
that in each case the CS entropy is positive and, in fact, equal to the
classical value.

\subsec
We begin by recalling the definition of the (classical) KS entropy.
Let $M$ be the phase space on which a probability measure $\nu$
and $\nu$-preserving automorphism $\varphi:M\rightarrow M$ are
defined. The latter means that $\varphi$ is a measurable bijective
function such that for all measurable sets $\cO$,
$\nu(\varphi(\cO))=\nu(\cO)$. Let $\cA=\{A_j\}$, $1\leq j\leq$, be
a finite partition of $M$ into measurable and pairwise
disjoint (up to measure zero) subsets. The entropy of this
partition is defined by
$$
H(\cA)=\sum_j\eta(\nu(A_j)),\ref{\infentref}
$$
where the function $\eta$ is given by
$$
\eta(t)=-t\log t,\qquad 0\leq t\leq 1.\ref{\etadefref}
$$
Clearly, $H$ is invariant under	$\varphi$,
$$
H(\varphi(\cA))=H(\cA),\ref{\invarianceref}
$$
where $\varphi(\cA)=\{\varphi(A_1),\ldots ,\varphi(A_n)\}$. Now,
given two such
partitions, $\cA$ and $\cB$, we form a finer partition $\cA\vee
\cB$ by taking the intersections of the elements of $\cA$ with
the elements of $\cB$. The entropy is subadditive with respect
to the operation $\vee$,
$$
H(\cA\vee\cB)\leq H(\cA)+H(\cB).\ref{\subadref}
$$
This and \invarianceref\ imply that the limit
$$
H(\cA ,\varphi)=\lim_{n\to\infty}{1\over n}H(\cA\vee \varphi(\cA)
\vee\ldots\vee \varphi^{n-1}(\cA))\num
$$
exists. The KS entropy of $\varphi$ is defined as the supremum of
$H(\cA,\varphi)$ over all possible choices of the finite partition $\cA$,
$$
h_{KS}(\varphi)=\sup_{\cA}H(\cA,\varphi).\ref{\ksentropyref}
$$
This definition does not lend itself to explicit computations.
However, the fundamental theorem of Kolmogorov and Sinai [CFS]
states that, in fact, $h_{KS}(\varphi)$ can be computed from a single
partition, provided that it is sufficiently generic. More precisely,
$h_{KS}(\varphi)=H(\cA,\varphi)$, if $\cA$ is a partition
such that the sets $\varphi^k(A_j),\, j=1,\ldots ,n\, ,\; k\in\bZ$,
generate the $\sigma$-algebra of measurable sets on $M$.

We will explain in Section \connessec\ how Connes and St\o rmer
generalized the theory outlined above to the non-commutative case.

\subsec
For later convenience we now briefly review the definitions
of the classical cat map and baker's map. For a more complete
presentation and a variety of results we refer the reader to [A],
[AW], and [CFS].

We consider an element $\gamma\in SL(2,\bZ)$,
$$
\gamma=\left(\matrix{a&b\cr
                c&d\cr}\right),\num
$$
with $|\tr(\gamma)|>2$. Such a matrix has two eigenvalues $\mu_1,
\mu_2$, with $\mu_1\mu_2=1$. We label them so that $|\mu_1|>1$,
and $|\mu_2|<1$. The action of $\gamma$ on the plane $\bR^2$ is given
as usual by $(x_1, x_2)\rightarrow (y_1, y_2)$ with
$$
\eqalign{
&y_1=ax_1+bx_2,\cr
&y_2=cx_1+dx_2.\cr
}\ref{\relinmapref}
$$
For later reference, we rewrite \relinmapref\ in terms of the
complex variable $z=(x_1+ix_2)/\sqrt{2}$ as $z\rightarrow w$, with
$$
w=\ol\alpha z +\beta\ol z,\ref{\complinmapref}
$$
where the complex parameters $\alpha$ and $\beta$ are given by
$$
\eqalign{
&\alpha=(a+d+i(b-c))/2,\cr
&\beta=(a-d+i(b+c))/2,\cr
}\ref{\alphabetaref}
$$
and satisfy $|\alpha|^2-|\beta|^2=1$. The transformation
\relinmapref\ is area preserving. Since the coefficients in
\relinmapref\ are integer, $\gamma$ also defines an area preserving
automorphism of the torus $\bT^2=\bR^2/\bZ^2$ which we will denote
by the same symbol $\gamma$. The group $\{\gamma^n\}_{n\in\bZ}$ of
automorphisms of $\bT^2$ is called the cat dynamics (in fact, this
is an example of Anosov dynamics).

The definitions above are of course meaningful without assuming
that $|\tr(\gamma)|>2$. The resulting dynamical systems are non-chaotic,
and, as such, less relevant to the subject of this paper.

It turns out that for the cat dynamics,
$$
h_{KS}(\gamma)=\log|\mu_1|,\ref{\catentref}
$$
where $\mu_1$ is the eigenvalue of $\gamma$ whose absolute value is
larger than $1$. A beautiful proof of this result in the context of
symbolic dynamics is presented in [AW]. If $|\tr(\gamma)|\leq 2$,
then $h_{KS}(\gamma)=0$, showing that the corresponding dynamics is
indeed non-chaotic.

The baker's map $B$ takes a
point $(x_1,x_2)$ of $\bT^2=\bR^2/\bZ^2$ to a point $(x_1^\prime,
x_2^\prime)$ of $\bT^2$ given by
$$
\eqalign{
x_1^\prime&=\cases{2x_1, &if $0\leq x_1<1/2$;\cr
                   2x_1-1, &if $1/2\leq x_1<1$,\cr}\cr
x_2^\prime&=\cases{x_2/2, &if $0\leq x_1<1/2$;\cr
                   (x_2+1)/2, &if $1/2\leq x_1<1$.\cr}\cr
}\ref\classbak
$$
The transformation $B$ is measure preserving. In order to prepare
ground for the quantization of $B$, we first rewrite \classbak\
in terms of generators of the algebra $L^\infty(\bT^2)$ of essentially
bounded functions on $\bT^2$. We set $g(x_1,x_2)=e^{2\pi ix_1},\;
h(x_1,x_2)=e^{2\pi ix_2}$. Then the transformation \classbak\ of $\bT^2$
is equivalent to the following automorphism of the algebra $L^\infty
(\bT^2)$ (which, for simplicity, is denoted by the same symbol $B$):
$$
\eqalign{
B(g)&=g^2,\cr
B(h)&=\sqrt{h}\big(2\chi_{[0,1/2)}(x_1)-1\big),\cr}
\ref{\classbaker}
$$
where the square root $\sqrt{h}$ is defined by $\sqrt{h}(x_1,x_2)=
e^{i\pi x_2}$, and where $\chi_{[0,1/2)}$ is the indicator function
of the interval $[0,1/2)$.

For the baker's map,
$$
h_{KS}(B)=\log 2.\ref{\bakentref}
$$

\subsec
One of the central concepts of this paper is that of quantization
of a dynamical system. Without getting involved with technicalities
we would like to emphasize several points which will explain the
particular conceptual framework which we chose to work with.

Quantization of a dynamical system has two components: kinematic
and dynamic. The kinematic component of quantization involves the
construction of a suitable quantized phase space of the system. This
quantized phase space is given in terms of a non-commutative algebra
$\fA_\hbar$ of observables. In the language of non-commutative
geometry, $\fA_\hbar$ is an algebra of functions on the quantized
phase space. Very much like in the classical situation, where
(depending on the problem) one might be interested in the study
of the algebra of continuous functions, smooth functions,
compactly supported smooth functions, measurable functions,
etc., specific choices of the composition of $\fA_\hbar$ can
be made. This may result in imposing the structure of a
$\bC^*$-algebra, a von Neumann algebra or some suitably defined
locally convex algebra, on the algebra of observables.

The dynamic component of quantization consists in defining a time
evolution on the quantized phase space. A natural way of doing
this is to find a suitable one parameter group of automorphisms of
$\fA_\hbar$, where the parameter (discrete or continuous) has the
meaning of time. Recall that an automorphism of an algebra $\fR$
is a linear one-to-one map $\Phi$ of $\fR$ onto itself such that
$\Phi(ab)=\Phi(a)\Phi(b)$, for all $a,b\in\fR$. If $\fR$ is an
algebra with involution, it is also required that $\Phi(a^*)=\Phi(a)^*$.

The ``suitability'' of the choices made, namely	that of the algebra
$\fA_\hbar$ and of the time evolution, is settled by the correspondence
principle. This amounts to showing that limits of the quantized objects,
as $\hbar\to 0$, yield the corresponding classical objects. Quantization
is a highly non-unique procedure, and the correspondence principle is
the only physical principle allowing one to decide whether a particular
procedure is correct. To our taste, the most satisfying mathematical
framework for quantization is that of ``strict deformation quantization''
proposed in [R1].

\subsec
Quantization of the  cat dynamics on the torus has been discussed
before by a number of authors. The original reference is [HB], where
a scheme is proposed using a group of unitary matrices on a finite
dimensional Hilbert space. The generator of this group was determined
from ($i$) the observation that the generating function of \relinmapref\
is quadratic, and ($ii$) the assumption that, in the quadratic case,
the semiclassical expressions are exact. This quantized dynamics was
further studied in [K1,2], [MO], [DGI], [BD], and [D], where a variety of
beautiful number theoretic results were derived.

A similar quantization scheme for baker's dynamics was first
proposed in [BV], and further refined and studied e.g. in [CTH],
[SV], [S], and [BDG]. These references are concerned with questions of
quantum chaology. The intrinsic simplicity of the baker's dynamics
has been very useful in studying these questions.

Our approach is slightly different, even though equivalent in the
sense specified at the end of previous subsection. It is based on
an infinite dimensional Hilbert space. The infinite dimensionality
of the Hilbert space is due to the occurrence of $\Theta$-vacua
(to use the language of quantum gauge field theory), which in turn
is a consequence of the fact that the phase space of the system,
namely the torus, is not simply connected. We study a non-abelian
algebra, known as the algebra of functions on a quantized torus
[R2], and identify a suitable group of automorphisms of this algebra
as the quantized dynamics.

\subsec
The paper is organized as follows. In Section II, we define
the quantized linear dynamics on the plane. This will be the
starting point for the construction of quantized cat
dynamics. In Section III, we review the construction of the
quantized torus, and show that the cat dynamics on the torus defines
a group of automorphisms of the quantized torus. This group
is the quantized cat dynamics on the torus. A construction
of quantized baker's dynamics is described in Section IV.
Section V has largely a review character. We explain the properties
and construction of the CS entropy, and establish a technical lemma.
Using this lemma, we compute, in Section VI, the CS entropy of the
quantized dynamics on the torus.

\section\planesec{Quantized linear dynamics on the plane}

\newsubsec
Of the many representations of quantum mechanics we choose
the Bargmann representation (see e.g. [F]), as in this representation
wave functions are defined on the phase space of the system.
It can also be generalized to phase spaces other than flat
spaces [B1], which should be important for future
extensions of the results of this paper. In the Bargmann
representation, the Hilbert space of states $\hil$ consists
of entire functions on $\bC$ which are square integrable with
respect to the probability measure $d\mu_\hbar(z)=
(\pi\hbar)^{-1}\exp\{-|z|^2/\hbar\}d^2z$. This Hilbert space
has two remarkable properties: ({\it i}) it has a reproducing
kernel, namely the function $\exp\{\ol wz/\hbar\}\in\hil$
satisfies the equation
$$
\int_{\bC}\exp\{\ol wz/\hbar\}\phi(w)d\mu_\hbar(w)=
\phi(z),\ref{\reprodref}
$$
for all $\phi\in\hil$, and ({\it ii}) it carries a unitary
projective representation of the group of translations of
$\bC$ given by
$$
U(\zeta)\phi(z)=\exp\big\{{1\over\hbar}(\ol\zeta z-|\zeta|^2/2)\big\}
\phi(z-\zeta),\quad \zeta\in\bC.\ref{\udefref}
$$
For future reference, we note that
$$
U(\zeta)U(\xi)=e^{i{\rm Im}(\ol\zeta\xi)/\hbar}U(\zeta+\xi).
\ref{\ucomref}
$$

The algebra of observables (or functions on the quantized plane)
can be defined as an algebra generated by Toeplitz operators. A
Toeplitz operator $T_\hbar(f)$ with symbol $f$ (where $f$ is a
measurable function on $\bC$) is defined by
$$
T_\hbar(f)\phi(z)=\int_{\bC}e^{z\ol w/\hbar}f(w)\phi(w)
d\mu_\hbar(w).\ref{\toepldefref}
$$
Various restrictions on the class of symbols $f$ may be imposed,
leading to various algebras of operators on $\hil$. Since the quantized
plane is not the main concern of this paper, we ignore this issue, and
refer the interested reader to e.g. [BC1,2] for precise statements.
See also [Z] for a related but more geometric approach.
For our needs, it is only important that all bounded continuous
functions are included in the class of symbols.

The Toeplitz operator with the symbol $f(z)=z$ is denoted by
$A^\dagger$, and the Toeplitz operator with $f(z)=\ol z$ is
denoted by $A$. These are the creation and annihilation operators
obeying the usual commutation relation
$$
[A,A^\dagger]=\hbar.\ref{\ccrref}
$$
In fact, the quantization map $f\rightarrow T_\hbar(f)$ can be
regarded as the anti-Wick ordering prescription, i.e., in the
quantized expressions, all the annihilation operators are placed
to the left of the creation operators.
\subsec
To a $\gamma$ as defined in the Introduction we assign the following
Bogolubov transformation $(A^\dagger, A)\rightarrow (B^\dagger,
B)$,
$$
\eqalign{
B^\dagger&=\ol\alpha A^\dagger+\beta A,\cr
B&=\alpha A+\ol\beta A^\dagger.\cr
}\ref{\bogref}
$$
We will now show that this transformation is unitarily implementable,
i.e. $B^\dagger=FA^\dagger F^{-1}$, and determine such a unitary
$F$ explicitly.

First, we note that the ground state $\omega_\gamma(z)$
for $B$ satisfies the differential equation:
$$
\hbar\alpha\omega_\gamma^\prime(z)+\ol\beta z\omega_\gamma(z)=0,\num
$$
and so
$$
\omega_\gamma(z)=|\alpha|^{-1/2}\exp\big\{-{{\ol\beta z^2}\over
{2\hbar\alpha}}\big\},\ref{\omegadefref}
$$
where the normalizing constant has been chosen so that $||\omega_
\gamma||=1$. We require that $F$ maps the function identically equal
$1$ (the ground state for $A$) to $\omega_\gamma$. Then, using the
Hausdorff-Baker-Campbell formula (see e.g. [F]),
$$
\eqalign{
F\exp\{\ol wz/\hbar\}
&=F\exp\{\ol wA^\dagger/\hbar\}F^{-1}\omega_\gamma(z)\cr
&=\exp\{\ol w(\ol\alpha A^\dagger+\beta A)/\hbar\}\omega_\gamma(z)\cr
&=\exp\{(\ol w\ol\alpha z+\ol\alpha\beta\ol w^2/2)/\hbar\}
\exp\{\ol w\beta {d\over{dz}}\}\omega_\gamma(z)\cr
&=\exp\{(\ol w\ol\alpha z+\ol\alpha\beta\ol w^2/2)/\hbar\}
\omega_\gamma(z+\ol w\beta)\cr
&=|\alpha|^{-1/2}\exp\{(\ol wz+\beta\ol w^2/2-\ol\beta z^2/2)
/\hbar\alpha\}.
}
$$
Using the fact that $\exp\{\ol wz/\hbar\}$ is the reproducing
kernel for the measure $d\mu_\hbar$ we thus find that the
action of $F$ on $\phi\in\hil$ is given by
$$
\eqalign{
F\phi(z)&=|\alpha|^{-1/2}\exp\big\{-{{\ol\beta z^2}\over
{2\hbar\alpha}}\big\}\int_{\bC}\exp\big\{{{\ol wz}\over
{\hbar\alpha}}+{{\beta\ol w^2}\over{2\hbar\alpha}}\big\}
\phi(w)d\mu_\hbar(w)\cr
&=T_\hbar(\omega_\gamma)S_{\gamma^{-1}}
T_\hbar(\omega_{\gamma^{-1}})^*\phi(z),\cr
}\ref{\fdefref}
$$
where $S_\gamma$ is defined by $S_\gamma\phi(z)=|\alpha|^{1/2}
\phi(z/\alpha)$. It is straightforward to verify that the inverse of
$F$ is given by
$$
\eqalign{
F^{-1}\phi(z)&=|\alpha|^{-1/2}\exp\big\{{{\ol\beta z^2}\over
{2\hbar\ol\alpha}}\big\}\int_{\bC}\exp\big\{{{\ol wz}\over
{\hbar\ol\alpha}}-{{\beta \ol w^2}\over{2\hbar\ol\alpha}}\big\}
\phi(w)d\mu_\hbar(w)\cr
&=T_\hbar(\omega_{\gamma^{-1}})S_\gamma
T_\hbar(\omega_\gamma)^*\phi(z),\cr
}\ref{\finvdefref}
$$
and that $F$ is unitary. Let us summarize the calculations above
in the following theorem.

\thm\fpropthm{There exists a unique unitary operator $F$
satisfying $FA^\dagger F^{-1}=B^{\dagger}$, and $F1=\omega_\gamma$.
This operator and its inverse are given by equations \fdefref\
and \finvdefref .}

\medskip
The group $\{F^n\}_{n\in\bZ}$ of unitary operators on $\hil$
is called the evolution group for the linear dynamics on the
plane. The corresponding group of automorphisms of the algebra
of observables is generated by $a\to FaF^{-1}$.

\subsec
There is a simple relation between $F$ and the unitary operators
$U(\zeta)$ defined in \udefref .

\thm\conjthm{The conjugation of $U(\zeta)$ by $F$ is equal to
$U(\gamma^{-1}\zeta)$,
$$
FU(\zeta)F^{-1}=U(\alpha\zeta-\beta\ol\zeta).\ref{\conjref}
$$}

\noindent{\it Proof.} The proof is a straightforward computation.
Using \fdefref\ and \finvdefref\ we find that
$$
\eqalign{
FU(\zeta)&F^{-1}\phi(z)=|\alpha|^{-1}\exp\big\{-{1\over{2\hbar}}
(|\zeta|^2+\ol\beta z^2/\alpha-\ol\beta\zeta^2/\ol\alpha)\big\}\cr
&\times\int_{\bC}\exp\big\{{1\over\hbar}\big({{\beta\ol w^2}\over
{2\alpha}}+{{\ol\beta w^2}\over{2\ol\alpha}}+{{z\ol w}\over
\alpha}+{{(\ol\alpha\ol\zeta-\ol\beta\zeta+\ol v)w}\over\alpha}
+{{\ol v\zeta}\over\alpha}+{{\beta\ol v^2}\over{2\ol\alpha}}\big)
\big\}\cr
&\qquad\times\phi(v)d\mu_\hbar(w)d\mu_\hbar(v).\cr
}
$$
Evaluating the $w$-integral and using $|\alpha|^2-|\beta|^2=1$ yields
$$
\eqalign{
FU(\zeta)F^{-1}\phi(z)&=\exp\big\{\big(z\ol{(\alpha\zeta-\beta
\ol\zeta)}-|\alpha\zeta-\beta\ol\zeta|^2/2\big)/\hbar\}\cr
&\times\int_{\bC}\exp\big\{(z-\alpha\zeta+\beta\ol\zeta)\ol v
/\hbar\big\}\phi(v)d\mu_\hbar(v),\cr
}
$$
which by means of \reprodref\ is equal to
$$
\exp\big\{z(\ol{\alpha\zeta-\beta\ol\zeta)}/\hbar-
|\alpha\zeta-\beta\ol\zeta|^2/2\hbar\big\}\phi(z-(\alpha\zeta-
\beta\ol\zeta))=U(\gamma^{-1}\zeta)\phi(z),
$$
as claimed. $\square$

\section\torussec{Quantized cat dynamics on the torus}

\newsubsec
Having defined the quantized linear dynamics on the plane
we now proceed to constructing the quantized cat dynamics
on the torus. As explained e.g. in [KL], one can regard the
quantized torus as a suitably defined quotient of the quantized
plane by the group $\bZ^2$. Namely, we define the algebra of
observables on the quantized torus to be the algebra of all
Toeplitz operators with continuous $\bZ^2$-invariant symbols.
Such symbols can be written as Fourier series, and so the algebra
of observables is generated by $T_\hbar(f_1)$ and $T_\hbar(f_2)$,
where $f_k(x_1, x_2)=\exp\{2\pi ix_k\}$. However, writing $ix_1=
i(z+\ol z)/\sqrt{2},\;\; ix_2=(z-\ol z)/\sqrt{2}$, we verify easily
that
$$
\eqalign{
&T(f_1)=e^{-\pi^2\hbar}U(-i\pi\hbar\sqrt{2}),\cr
&T(f_2)=e^{-\pi^2\hbar}U(\pi\hbar\sqrt{2}).\cr
}\num
$$
It is thus natural to set
$$
\eqalign{
&U=U(-i\hbar\pi\sqrt{2}),\cr
&V=U(\hbar\pi\sqrt{2}),\cr
}\ref{\uvdefref}
$$
and regard the operators $U$ and $V$ as generators of the algebra
of functions on the quantized torus. Commutation relation \ucomref\
implies	that they obey the following set of relations:
$$
\eqalign{
&UU^*=U^*U=I,\cr
&VV^*=V^*V=I,\cr
&UV=e^{i\lambda}VU,\cr
}\ref{\relref}
$$
where for convenience we set $\lambda=4\pi^2\hbar$.
The algebra generated by $U$ and $V$ with the relations above has been
studied extensively by both physicists and mathematicians, and we
refer the reader to [R2] for an overview and extensive list of references.
In particular, it has been established that ``smooth elements'' in this
algebra obey a strong version of the correspondence principle [R1].

\subsec
For our purposes, we consider the von Neumann algebra $\fA_\hbar$, generated
by $U$ and $V$. Recall [D2] that an algebra of bounded operators $\fR$ on
a Hilbert space $\cH$ is called a von Neumann algebra, if ($i$) it
is closed under taking the hermitian conjugate, and ($ii$) it is equal
to its bicommutant, $\fR=\fR^{\prime\prime}$. Here $\fR^{\prime\prime}
=(\fR^\prime)^\prime$, where the commutant $\cS^\prime$ of a set of
operators $\cS$ on $\cH$ is defined as the set of all bounded operators
on $\cH$ which commute with all the elements of $\cS$. The von Neumann
algebra generated by a set $\cS$ is defined as the smallest von Neumann
algebra containing $\cS$. If $\cS$ is closed under taking the hermitian
adjoint, this turns out to be $\cS^{\prime\prime}$. In other words,
$\fA_\hbar=\{U,	U^*, V, V^*\}^{\prime\prime}$. In fact, $\fA_\hbar$
is isomorphic to the universal enveloping von Neumann algebra generated
by $U$ and $V$ which obey the relations \relref . This means, in
particular, that \relref\ are the only relations between $U$ and $V$.
One can think of the elements of $\fA_\hbar$ as bounded (but not
necessarily continuous) functions on the quantized torus.

The von Neumann algebra $\fA_\hbar$ is hyperfinite (i.e. it is a
closure of an increasing subsequence of finite dimensional subalgebras)
and can be equipped with a finite faithful trace. We will not reproduce
here the precise definitions (see e.g. [D2]). One should just keep in mind
a typical example, that of an algebra $L^\infty(M)$ of essentially bounded
functions on a compact space $M$ with a Borel probability measure $d\nu$.
Such a trace is then given by
$$
\tau(f)=\int_Mf\; d\nu.\ref{\tauabeldef}
$$
On the algebra $\fA_\hbar$, a faithful normal trace is determined by
$$
\tau_\hbar(\sum_{j,k}\alpha_{jk}U^jV^k)=\alpha_{00}.\ref{\tauhbardef}
$$

\subsec
Let us now derive the transformation rules for $U$ and $V$ under
conjugation by the operator $F$. Using \conjthm\ and \ucomref\
we obtain
$$
\eqalign{
FUF^{-1}&=U(-i\hbar\pi\sqrt{2}(\alpha+\beta))\cr
        &=U(-i\hbar\pi(a+ib)\sqrt{2})\cr
        &=e^{-i\hbar\pi^2ab}U(-i\hbar\pi\sqrt{2} a)U(\hbar\pi\sqrt{2} b)\cr
        &=e^{-i\lambda ab/2}U^aV^b,\cr
}
$$
and likewise
$$
FVF^{-1}=e^{-i\lambda cd/2}U^cV^d.
$$
These expressions define an automorphism $\Gamma_\hbar$ of $\fA_\hbar$.
We call the group $\{\Gamma_\hbar^n\}_{n\in\bZ}$ of automorphisms
generated by $\Gamma_\hbar$ the quantized cat dynamics on the torus.

\thm\anosautthm{The transformation
$$
\eqalign{
&\Gamma_\hbar(U)=e^{-i\lambda ab/2}U^aV^b,\cr
&\Gamma_\hbar(V)=e^{-i\lambda cd/2}U^cV^d.\cr
}\ref{\tref}
$$
defines an automorphism of $\fA_\hbar$.The trace $\tau_\hbar$ is
invariant under $\Gamma_\hbar$, i.e. $\tau_\hbar(\Gamma_\hbar(a))=
\tau_\hbar(a)$.}

\medskip\noindent{\it Proof.}
We need to show that $\Gamma_\hbar(U)$ and $\Gamma_\hbar(V)$ form a new
set of generators. Using \relref\ we compute:
$$
\Gamma_\hbar(U)^*=e^{i\lambda ab/2}V^{-b}U^{-a}=
\big(e^-{i\lambda ab/2}U^aV^b\big)^{-1}=\Gamma_\hbar(U)^{-1}.
$$
Likewise, $\Gamma_\hbar(V)^*=\Gamma_\hbar(V)^{-1}$. Furthermore,
$$
\eqalign{
\Gamma_\hbar(U)\Gamma_\hbar(V)
&=e^{-i\lambda(ab+cd)/2}U^aV^bU^cV^d\cr
&=e^{-i\lambda(ab+cd)/2-i\lambda bc}U^cU^aV^bV^d\cr
&=e^{-i\lambda(ab+cd)/2+i\lambda (ad-bc)}U^cV^dU^aV^b\cr
&=e^{-i\lambda}\Gamma_\hbar(V)\Gamma_\hbar(U).\cr
}
$$
We also note that the inverse of $\Gamma_\hbar$ is given by
$$
\eqalign{
&\Gamma_\hbar^{-1}(U)=e^{i\lambda bd/2}U^dV^{-b},\cr
&\Gamma_\hbar^{-1}(V)=e^{i\lambda ac/2}U^{-c}V^a.\cr
}\ref{\tinvref}
$$
Finally, the $\Gamma_\hbar$-invariance is an immediate consequence
of \tauhbardef . $\square$

\subsec
At this point it is not quite clear that $\Gamma_\hbar$ is indeed
a quantization of the classical map $\gamma$, i.e. that its classical
limit $\hbar\to 0$ indeed yields $\gamma$. The goal of this subsection
is to show that it is so. We let $||\cdot||_\hbar$ denote the operator
norm on the Hilbert space $\hil$.

\thm\deformthm{Let f be a continuous $\bZ^2$-invariant function on
$\bC$. Then:
$$
||FT_\hbar(f)F^{-1}-T_\hbar(f\circ\gamma)||_\hbar\to 0,\quad{\rm as}
\quad\hbar\to 0.\ref{\classlimref}
$$}

\noindent{\it Proof.}
Let $\epsilon>0$. We are going to show that for all sufficiently
small $\hbar$,
$$
||FT_\hbar(f)F^{-1}-T_\hbar(f\circ\gamma)||_\hbar\leq \epsilon .
\ref{\epsestref}
$$
We proceed in steps.

\noindent{\it Step 1.} By the Stone-Weierstra\ss\ theorem, there is
a trigonometric polynomial $P$ such that
$$
||f-P||_\infty\leq \epsilon /3,
$$
where $||f||_\infty=\sup_z|f(z)|$ is the usual sup-norm. Since
the operator norm of a Toeplitz operator does not exceed the
sup-norm of its symbol (see e.g. [B1]), $||T_\hbar(f)||_\hbar\leq||f||_\infty$,
this yields the following inequality:
$$
||T_\hbar(f)-T_\hbar(P)||_\hbar\leq\epsilon /3.\ref{\epsoneref}
$$

\noindent{\it Step 2.}
A trigonometric polynomial $P(z)$ is a linear combination of
terms of the form $\exp(\ol wz-\ol zw)$. In terms of the
creation and annihilation operators, for the corresponding
Toeplitz operator we have:
$$
T_\hbar(e^{\bar wz - \ol zw})=e^{-wA}e^{\ol wA^\dagger}.
$$
Conjugating the above equation by $F$ yields:
$$
\eqalign{
FT_\hbar(e^{\ol wz - \ol zw})F^{-1}&=Fe^{-wA}F^{-1}
Fe^{\ol wA^\dagger}F^{-1}\cr
&=e^{-w(\alpha A +\ol\beta A^{\dagger})}
e^{\ol w(\ol\alpha A^\dagger+\beta A)}\cr
&=e^{-\hbar(\alpha\ol\beta w^2+\ol\alpha\beta\ol w^2)/2}
e^{-\alpha wA}e^{-\ol\beta wA^\dagger}
e^{\beta \ol wA}e^{\ol\alpha\ol wA^\dagger},\cr
}
$$
where we have used the Hausdorff-Baker-Campbell formula as in the
derivation of \fdefref . Commuting the third and the fourth terms
gives further:
$$
\eqalign{
FT_\hbar(e^{\ol wz - \ol zw})F^{-1}&=
e^{-\hbar (\alpha\ol\beta w^2+\ol\alpha\beta\ol w^2 -
2|\beta|^2|w|^2)/2}e^{-w\alpha A +\ol w\beta A}
e^{\ol w\ol\alpha A^\dagger-w\ol\beta A^{\dagger}}\cr
&=e^{-\hbar (\alpha\ol\beta w^2+\ol\alpha\beta\ol w^2 -
2|\beta|^2|w|^2)/2}T_\hbar(e^{-w\alpha \ol z +\ol w\beta \ol z+
\ol w\ol\alpha z-w\ol\beta z})\cr
&=e^{-\hbar (\alpha\ol\beta w^2+\ol\alpha\beta\ol w^2 -
2|\beta|^2|w|^2)/2}T_\hbar(e^{\ol w(\ol\alpha z+\beta\ol z)-
w(\alpha\ol z+\ol\beta z)})\cr
&=e^{-\hbar (\alpha\ol\beta w^2+\ol\alpha\beta\ol w^2-2|\beta|^2
|w|^2)/2}T_\hbar(e^{{\ol w\gamma(z)-\ol{\gamma(z)}w }}).\cr
}
$$
We can thus make the following estimate:
$$
\eqalign{
||FT_\hbar(e^{\ol wz - \ol zw})&F^{-1}-T_\hbar(e^{{\ol w\gamma(z) -
\ol{\gamma(z)}w }})||_\hbar\cr
&\leq |e^{-\hbar (\alpha\ol\beta w^2+\ol\alpha\beta\ol w^2 -
2|\beta|^2|w|^2)/2}-1|\,||T_\hbar(e^{{\ol w\gamma(z)-\ol{\gamma(z)}w}})
||_\hbar\cr
&\leq |e^{-\hbar (\alpha\ol\beta w^2+\ol\alpha\beta\ol w^2 -
2|\beta|^2|w|^2)/2}-1|.\cr
}
$$
Clearly, the right hand side of the above inequality goes to zero,
as $\hbar\to 0$. Since $P$ is a linear combination of finitely many
terms of the above form, we can find $\delta$ (depending on $P$)
such that for $\hbar <\delta$ we have:
$$
||FT_\hbar(P)F^{-1}-T_\hbar(P\circ\gamma)||_\hbar\leq\epsilon /3.
\ref{\epstworef}
$$

\noindent{\it Step 3.}
We can now conclude the argument:
$$
\eqalign{
||FT_\hbar(f)F^{-1}-&T_\hbar(f\circ\gamma)||_\hbar\cr
&\leq||FT_\hbar(f)F^{-1}-FT_\hbar(P)F^{-1}||_\hbar +
||FT_\hbar(P)F^{-1}-T_\hbar(P\circ\gamma)||_\hbar\cr
&\quad +||T_\hbar(P\circ\gamma)-T_\hbar(f\circ\gamma)||_\hbar\cr
&\leq||T_\hbar(f)-T_\hbar(P)||_\hbar +
\epsilon/3+
||P\circ\gamma - f\circ\gamma||_\infty\cr
&\leq 2||f-P||_\infty +\epsilon/3\cr
&\leq\epsilon,\cr
}
$$
where we have used \epsoneref\ and \epstworef . $\square$

\subsec
So far the value of Planck's constant has not been restricted
in any way other than it should be a positive number. In particular,
the von Neumann algebra $\fA_\hbar$ is a well defined object
for all such $\hbar$. On the other hand, its structure depends
crucially on whether $\lambda/2\pi$ is a rational number or
not. It is well known that physics requires $\lambda/2\pi$ to
be rational. The standard informal argument, going back to Planck,
is that the volume of the phase space should be an integer multiple
of the elementary cell volume $2\pi\hbar$. Hence
$$
\hbar={1\over{2\pi N}},\qquad N\in\bN,\ref{\hbarrestref}
$$
or
$$
\lambda={2\pi\over N}.\ref{\lambdarestref}
$$
Incidentally, this is precisely the integrality condition
of geometric quantization which requires the symplectic form
on the torus divided by $2\pi\hbar$ to define a deRham
cohomology class with integer coefficients. Throughout the
rest of this paper, we will be assuming that the condition
above is satisfied. Trivial changes in our arguments
show that the conclusions below hold for arbitrary
positive rational $\lambda/2\pi$.

\subsec
The von Neumann algebra $\fA_\hbar$ has a simple structure
which is described in the theorem below. This theorem is
well known, and the references to the original literature
can be found in [R2]. Since the proof is not easy to extract
from the original references (and for the sake of completeness),
we include an elementary proof.	We denote by $\cM_N$ the (von
Neumann) algebra of complex $N\times N$ matrices, while by
$L^\infty(\bT^2)$ we denote the space of all essentially
bounded functions on the torus regarded as a von Neumann
algebra on the Hilbert space $L^2(\bT^2)$.

\thm\tensprodthm{We have the following isomorphism of von
Neumann algebras
$$
\iota:\quad\fA_\hbar\rightarrow L^\infty(\bT^2)\otimes\cM_N.
\ref{\factref}
$$
Under this isomorphism, the trace $\tau_\hbar$ factorizes into
a tensor product of traces,
$$
\tau_\hbar\circ\iota^{-1}=\tau\otimes(1/N)\;{\rm tr},\ref{\trfactref}
$$
where $\tau$ is given by \tauabeldef .}

\medskip\noindent{\it Proof.} It is clear from the relations \relref\
that $U^N$ and $V^N$ are in the center of $\fA_\hbar$. Let us
denote by $\fZ$ the von Neumann algebra generated by
$$
X=U^N,\quad\hbox{and } Y=V^N.\ref{\utonref}
$$
Obviously, $\fZ$ is isomorphic with $L^\infty(\bT^2)$,
with the isomorphism given by $X\rightarrow e^{2\pi
i\theta_1}$ and $Y\rightarrow	e^{2\pi i\theta_2}$. Consider
now the following (discontinuous) functions in $L^\infty(\bT^2)$:
$f_1(\theta)=e^{2\pi i\theta_1/N}$ and $f_2(\theta)=e^{2\pi
i\theta_2/N}$, and let $Z_1$ and $Z_2$ be the corresponding
elements of $\fZ$. Then the two elements $u=Z_1^{-1}U$
and $v=Z_2^{-1}V$ obey the following set of relations:
$$
\eqalign{
&uu^*=u^*u=I,\cr
&vv^*=v^*v=I,\cr
&uv=e^{i\lambda}vu,\cr
&u^N=v^N=I.\cr
}\ref{\reltworef}
$$
This algebra has the following realization. In the Hilbert space
$\bC^N$, choose an orthonormal basis $e_1,\ldots ,e_N$, and set
$ue_j=e^{i(j-1)\lambda}e_j,\;\; ve_j=e_{j+1}$, where $e_{N+1}=
e_1$ (by a slight abuse of notation, we denote the matrix
representatives of $u$ and $v$ by the same symbols). A	short
computation shows that the only matrices commuting
with $u$ and $v$ are scalar multiples of the identity, and thus
the von Neumann algebra generated by $u$ and $v$ can be identified
with the full matrix algebra $\cM_N$.

We have $U=Z_1u$, $V=Z_2v$, and the required isomorphism is given
by
$$
\iota(U)=f_1\otimes u,\quad\iota(V)=f_2\otimes v.\qquad
\ref{\isoref}
$$

To prove \trfactref , we note that
$$
\big(\tau\otimes(1/N)\;{\rm tr}\big)(f_1^jf_2^k\otimes u^jv^k)=
\int_0^1\int_0^1e^{2\pi i(j\theta_1+k\theta_2)/N}d\theta_1
d\theta_2\;(1/N)\;{\rm tr}(u^jv^k).\ref{\trcompref}
$$
Using the explicit realization of the operators $u$ and $v$ we
see that ${\rm tr}(u^jv^k)=0$, unless $k=pN,\; p\in\bZ$. However,
$\int_0^1e^{2\pi ip\theta_2}d\theta_2=0$, for $p\neq 0$, and so
\trcompref\ is zero for $k\neq 0$. Let $k=0$, and $j=Np+q,\;
0\leq q\leq N-1$. If $q>0$, then ${\rm tr}(u^j)=0$. If $q=0$, but
$p\neq 0$, then $\int_0^1e^{2\pi ip\theta_1}d\theta_1=0$. Consequently,
$$
\big(\tau\otimes(1/N)\;{\rm tr}\big)(f_1^jf_2^k\otimes u^jv^k)=
\delta_{j0}\delta_{k0}=\tau_\hbar(U^jV^k),\num
$$
and the claim follows. $\square$

Let us parenthetically remark that the corresponding result for
the $\bC^*$-algebra of functions on a quantized torus involves
a bundle of full matrix algebras over the torus rather than a
tensor product [R2].

\subsec
It is now easy to see that, under the isomorphism above, the
automorphism $\Gamma_\hbar$ becomes a tensor product of automorphisms of
the factors in \isoref .

\lemma\scalaroplemma{For $f\in L^\infty(\bT^2)$,
$$
\iota\Gamma_\hbar\iota^{-1} (f(\theta)\otimes I) =f(\gamma\theta+
\Delta_\gamma)\otimes I,\ref{\scalarpart}
$$
where
$$
\Delta_\gamma=(Nab/2,Ncd/2)
$$
is a constant.}

\medskip\noindent
{\it Proof.} Expanding $f$ in a Fourier series and using \utonref ,
we can write
$$
\iota^{-1}(f\otimes I)=\sum_{m,n\in\bZ}\widehat f_{m,n}X^m
Y^n=\sum_{m,n\in\bZ}\widehat f_{m,n}U^{Nm}V^{Nn},
$$
and thus
$$
\eqalign{
\Gamma_\hbar\iota^{-1}(f\otimes I)&=\sum_{m,n\in\bZ}\widehat f_{m,n}
(e^{-\pi iab/N}U^aV^b)^{Nm}(e^{-\pi icd/N}U^cV^d)^{Nn}\cr
&=\sum_{m,n\in\bZ}\widehat f_{m,n}(e^{\pi iNab}U^{Na}
V^{Nb})^m(e^{\pi iNcd}U^{Nc}V^{Nd})^n\cr
&=\sum_{m,n\in\bZ}\widehat f_{m,n}(e^{\pi iNab}X^a
Y^b)^m(e^{\pi iNcd}X^cY^d)^n,\cr}
$$
and the claim follows. $\square$

\thm\autfactthm{We have
$$
\iota \Gamma_\hbar\iota^{-1} = \Psi_\hbar\otimes\Phi_\hbar ,\num
$$
where $\Psi_\hbar$ is an automorphism of $L^\infty(\bT^2)$ given by
$$
\eqalign{
&\Psi_\hbar(e^{2\pi i\theta_1})=e^{2\pi i(a\theta_1+b\theta_2+Nab/2)},\cr
&\Psi_\hbar(e^{2\pi i\theta_2})=e^{2\pi i(c\theta_1+d\theta_2+Ncd/2)},\cr
}\ref{\psiref}
$$
and where $\Phi_\hbar$ is an automorphism of $\cM_N$ given by
$$
\eqalign{
&\Phi_\hbar(u)=e^{-i\lambda(N+1)ab/2}u^av^b,\cr
&\Phi_\hbar(v)=e^{-i\lambda(N+1)cd/2}u^cv^d.\cr
}\ref{\phiref}
$$
}

\noindent
Notice that in the case when $ab$ and $cd$ are even (this case is
referred to as ``quantizable'' in [HB]) $\Psi_\hbar$ coincides with
the classical map \relinmapref . It is thus natural to regard
$\Psi_\hbar$ as the classical component of the dynamics, and
$\Phi_\hbar$ its purely quantum component.

\medskip\noindent
{\it Proof.} The algebra $L^\infty(\bT^2)\otimes\cM_N$ is generated
by elements of the form $f\otimes u$ and $f\otimes v$. In view of
\scalaroplemma , it is sufficient to compute $\iota\Gamma_\hbar\iota^{-1}
(I\otimes u)$ and $\iota\Gamma_\hbar\iota^{-1}(I\otimes v)$. Using the
notation introduced in the proof of \tensprodthm , we have
$$
\eqalign{
\Gamma_\hbar\iota^{-1}(I\otimes u)&=\Gamma_\hbar(Z_1^{-1})\Gamma_\hbar(U)\cr
&=e^{-i\lambda ab/2-i\lambda Nab/2}U^aV^bZ_1^{-a}Z_2^{-b}\cr
&=e^{-i\lambda(N+1)ab/2}u^av^b.\cr}
$$
The calculation for $I\otimes v$ is analogous. $\square$

\section\bakersec{Quantized baker's maps}

\newsubsec
In this section we introduce a group of automorphisms of $\fA_\hbar$
which we call the quantized baker's dynamics. Our construction
requires that $N$ in \lambdarestref\ be an odd number, and we
make this assumption throughout the section. This is unlike the
quantization procedure proposed in [BV], [CTH], [SV], [S], and
[BDG], which requires $N$ to be even. We do not know yet whether our
quantization is equivalent to it. Because of its
discontinuous character, the quantized baker's dynamics can
be defined in the framework of von Neumann algebras only. This
should be contrasted with the cat dynamics, where we chose to
work with von Neumann algebras rather than $\bC^*$-algebras
for the reason of convenience only.

First, we review some facts from operator calculus. If $S$ is a unitary
operator, then by $E_S(\sigma)$ we will denote its spectral measure.
In other words, $S=\int_0^1e^{2\pi i\sigma}dE_S(\sigma)$. For any
real number $\alpha$, we define $S^\alpha=\int_0^1e^{2\pi i\alpha
\sigma}dE_S(\sigma)$ (in particular, $S^{1/2}=\int_0^1e^{\pi i\sigma}
dE_S(\sigma)$). It follows by functional calculus that $S^\alpha$
is unitary, and so $S^\alpha=\int_0^1e^{2\pi i\sigma}dE_{S^\alpha}
(\sigma)$. It is easy to express the spectral measure $E_{S^\alpha}$
in terms of $E_S$. In particular,
$$
E_{S^n}(\sigma)=\sum_{0\leq j\leq n-1}E_S\big({{\sigma+j}\over n}\big)-
E_S\big({j\over n}\big),\quad\hbox{for}\;\;n\in\bN,\ref{\esnref}
$$
$$
E_{S^{1/2}}(\sigma)=\cases{E_S(2\sigma), &if $0\leq\sigma<1/2$;\cr
                           I, &if $1/2\leq\sigma<1$,\cr}\ref{\eonehalfref}
$$
and
$$
E_{S^{-1}}(\sigma)=E_S(1-\sigma).\ref{\einvref}
$$
Obviously, $(S^{1/2})^2=S$. However, $(S^2)^{1/2}\neq S$. The latter
fact will play a role in the following, and we state it as a lemma.
\lemma\sqrtlemma{Let $S$ be unitary. Then
$$
(S^2)^{1/2}=S\;\big(2E_S(1/2)-I\big).\ref{\ssquareroot}
$$}
\noindent
{\it Proof.} We use \esnref\ to compute:
$$
\eqalign{
(S^2)^{1/2}&=\int_0^1e^{i\pi\sigma}dE_{S^2}(\sigma)\cr
           &=\int_0^1e^{i\pi\sigma}dE_S\big({\sigma\over 2}\big)
	    +\int_0^1e^{i\pi\sigma}dE_S\big({\sigma+1\over 2}\big)\cr
           &=\int_0^{1/2}e^{2\pi i\sigma}dE_S(\sigma)
	    -\int_{1/2}^1e^{2\pi i\sigma}dE_S(\sigma)\cr
	   &=SE_S(1/2)-S\;\big(I-E_S(1/2)\big). \qquad\square\cr}
$$

\subsec
We now come back to the algebra \relref . For a unitary
$S\in\fA_\hbar$ we define
$$
\eqalign{
\sqrt{S}&=(S^{-N})^{1/2}S^{(N+1)/2},\cr
P(S)&=E_{S^N}(1/2).\cr}\ref{\defsref}
$$
Clearly, $\sqrt{S}$ is a particular square root of $S$,
$$
(\sqrt{S})^2=S.\ref{\squareref}
$$
Furthermore,
$$
(\sqrt{S})^N = (S^N)^{1/2}.\ref{\nthpowerref}
$$
Note also that since $N$ is odd and $V^N$ is central, the following
commutation relation between $U$ and $\sqrt{V}$ holds:
$$
U\sqrt{V}=-e^{i\lambda/2}\sqrt{V}U.\ref{\urootv}
$$

Consider now the following transformation on the generators of
$\fA_\hbar$:
$$
\eqalign{
\bak(U)&=U^2,\cr
\bak(V)&=\sqrt{V}\big(2P(U)-I\big).\cr
}\ref\bakerdef
$$
We extend $\bak$ to $\fA_\hbar$ by requiring that $\bak(ab)
=\bak(a)\bak(b)$ and $\bak(a^*)=\bak(a)^*$.

\thm\automorphthm{The transformation $\bak$ defines a
$\tau_\hbar$-preserving $*$-automorphism of the von Neumann algebra
$\fA_\hbar$.}

\medskip\noindent
{\it Proof.} We need to verify that $\bak(U)$ and $\bak(V)$ obey
the same relations as $U$ and $V$, and that $\bak$ has an inverse.
The former property is an immediate consequence of \urootv , while
the latter one can be established as follows. Let $T$ be a
$*$-antiautomorphism of $\fA_\hbar$ defined by $T(U)=V$ and $T(V)=U$
(clearly, $T$ preserves	\relref , as $T(UV)=T(V)T(U)$). Consider
now the $*$-automorphism $T\bak T$. Using the fact that
$$
\bak(V)^2=\big(\sqrt{V}(2P(U)-I)\big)^2=V,\num
$$
we immediately find that
$$
\eqalign{
(T\bak T)\bak(U)&=T\bak T(U^2)=T\bak(V^2)=T\bak(V)^2=T(V)=U,\cr
\bak(T\bak T)(V)&=\bak T\bak(U)=\bak T(U^2)=\bak(V^2)=V.\cr
}
$$
It is slightly more difficult to verify the remaining two relations.
We have:
$$
\eqalign{
(T\bak T)\bak(V)&=T\bak T\big(\sqrt{V}(2P(U)-I))\big)
=T\bak\big(\sqrt{U}(2P(V)-I))\big)\cr
&=T\big((\sqrt{U^2}(2P(\bak(V))-I))\big).\cr}
$$
Now, according to \sqrtlemma\ and \einvref ,
$$
\eqalign{
\sqrt{U^2}&=(U^{-2N})^{1/2}U^{N+1}=U\;\big(2E_{U^{-N}}(1/2)-I\big)\cr
          &=U\;\big(2E_{U^N}(1/2)-I\big)=U\;\big(2P(U)-I\big).\cr
}\num
$$
Furthermore, using \eonehalfref\ and \nthpowerref ,
$$
\eqalign{
P(\bak(V))&=E_{\bak(V)^N}(1/2)=E_{(V^N)^{1/2}(2P(U)-I)}(1/2)\cr
          &=E_{(V^N)^{1/2}}(1/2)P(U)+\big(I-E_{(V^N)^{1/2}}(1/2)\big)
            \big(I-P(U)\big)\cr
          &=P(U),\cr}
$$
and so
$$
(T\bak T)\bak(V)=T\big(U\,(2P(U)-I)^2\big)=T(U)=V.
$$
In the same fashion we verify the last relation:
$$
\eqalign{
\bak(T\bak T)(U)&=\bak T\bak(V)=\bak T\big(\sqrt{V}(2P(U)-I))\big)
=\bak\big(\sqrt{U}(2P(V)-I))\big)\cr
&=\sqrt{U^2}\big(2P(\bak(V))-I\big)=U,\cr
}
$$
and so $T\bak T=\bak^{-1}$.

The $\tau_\hbar$-invariance of $\bak$ can be easily verified by
means of \trfactref\ and the next theorem.
$\square$

The automorphism $\bak$ of $\fA_\hbar$ is called the quantized
baker's map.

\subsec
The fundamental property of $\bak$ is that it factorizes under the
isomorphism \factref .

\thm\factorthm{We have
$$
\iota \Gamma_\hbar\iota^{-1} = \Psi\otimes\Phi_\hbar ,\num
$$
where $\Psi$ is an automorphism of $L^\infty(\bT^2)$ given by
$$
\eqalign{
&\Psi(e^{2\pi i\theta_1})=e^{4\pi i\theta_1},\cr
&\Psi(e^{2\pi i\theta_2})=e^{\pi i\theta_2}
\big(2\chi_{[0,1/2)}(\theta_1)-1\big),\cr
}\ref{\psiref}
$$
and where $\Phi_\hbar$ is an automorphism of $\cM_N$ given by
$$
\eqalign{
&\Phi_\hbar(u)=u^2,\cr
&\Phi_\hbar(v)=v^{(N+1)/2}.\cr
}\ref{\phiref}
$$}

\noindent
Observe that $\Psi$ coincides with \classbaker . As in the case of
the cat dynamics, one can think about $\Psi$ as the purely classical
component of the dynamics, and about $\Phi_\hbar$ as its purely quantum
component.

\medskip\noindent
{\it Proof.} Proceeding as in the proof of \scalaroplemma , we
readily find that
$$
\iota B_\hbar\iota^{-1}(f\otimes I)=Bf\otimes I.\num
$$
Furthermore,
$$
\iota\bak\iota^{-1}\big(e^{2\pi i\theta_1/N}\;\otimes\; u\big)
=\iota\bak(U)=\iota(U^2)
=e^{4\pi i\theta_1/N}\;\otimes\; u^2.
$$
Similarly,
$$
\eqalign{
\iota\bak\iota^{-1}\big(e^{2\pi i\theta_2/N}\;\otimes\; v\big)
&=\iota\bak(V)=\iota\big((V^{-N})^{1/2}V^{(N+1)/2}(2P(U)-I)\big)\cr
&=e^{i\pi\theta_2/N}\big(2\chi_{[0,1/2)}(\theta_1)-1\big)
\otimes v^{(N+1)/2},\cr}
$$
and the proof is complete. $\square$

\section\connessec{Connes-St\o rmer entropy}

\newsubsec
To motivate the construction of the CS entropy we first reformulate
the definition of the classical KS entropy in purely algebraic (rather
than measure theoretic)	terms (see also [B3]). We assume that $M$ is a
compact phase space with a Borel probability measure $d\nu$ defined on
it, and	$\tau$ define the faithful normal trace on $L^\infty(M)$ given
by \tauabeldef . Given a partition $\cA$ of $M$ (defined as
in the Introduction), we consider the finite dimensional subalgebra
$\fN\subset L^\infty(M)$ which is generated by the characteristic
functions $\chi_{A_j}$. The operator of multiplication by $\chi_{A_j}$
is a projection operator and we denote it by $p_j$. Note that each
projection $p_j$ is minimal (i.e. is not a sum of two non-trivial
projections in $\fN$), and $\sum_jp_j=I$. We define the entropy of
the subalgebra $\fN$ to be
$$
H(\fN)=\sum_j\tau(\eta(p_j))=H(\cA).\num
$$
For two such subalgebras, $\fN_1$ and $\fN_2$, we let $\fN_1\vee\fN_2$
denote the (finite dimensional)	subalgebra generated by $\fN_1$ and
$\fN_2$.

Now, a measure preserving automorphism $\varphi$ of $M$ defines a
$\tau$-preserving automorphism $\Phi$ of $\fR$,
$$
\Phi f(x)=f\circ\varphi(x).\ref{\Phidefref}
$$
We set
$$
H(\fN,\Phi)=\lim_{k\to\infty}{1\over k}H(\fN\vee\Phi(\fN)\vee
\ldots\vee\Phi^{k-1}(\fN))=H(\cA,\varphi),
$$
and define the entropy of the automorphism $\Phi$ as the supremum
of this quantity over all possible choices of $\fN$ (this is, of
course, equal to $h_{KS}(\varphi)$).
\subsec
The construction above of the entropy of a measure preserving
automorphism was generalized to the non-commutative case by
Connes and St\o rmer [CS] (in the von Neumann algebraic setup),
and later by Connes, Narnhofer and Thirring [CNT] (in the
$\bC^*$-algebraic setup). We choose the original Connes-St\o rmer
construction as it suits our needs best.

Let $\fR$ be a von Neumann algebra, and let $\tau$ be a
finite faithful normal trace on $\fR$. Consider	a collection
$\fN_1,\ldots ,\fN_k$ of finite dimensional von Neumann subalgebras
of $\fR$. The key difficulty to overcome here is the fact that
$\fN\vee\fP$ may not be finite dimensional, even though
$\fN$ and $\fP$ are. Connes and St\o rmer defined a function
$H(\fN_1,\ldots ,\fN_k)$ which replaces $H(\fN_1\vee\ldots\vee\fN_k)$
but reduces to it in the commutative case. Specifically, this
function satisfies the following properties:

\item{(A)} $H(\fN_1,\ldots ,\fN_k)\leq H(\fP_1,\ldots ,\fP_k)$,
if $\fN_j\subset\fP_j$, for all $1\leq j\leq k$;
\item{(B)} $H(\fN_1,\ldots ,\fN_m,\fN_{m+1},\ldots ,\fN_n)\leq
H(\fN_1,\ldots ,\fN_m)+H(\fN_{m+1},\ldots ,\fN_n)$;
\item{(C)} if $\fN_1,\ldots ,\fN_m\subset\fN$, then $H(\fN_1,\ldots ,
\fN_m, \fN_{m+1},\ldots ,\fN_n)\leq H(\fN,\fN_{m+1},\ldots ,\fN_n)$;
\item{(D)} if $\{p_\alpha\}$ is any family of minimal projections
in $\fN$ such that $\sum_\alpha p_\alpha=I$, then $H(\fN)=
\sum_\alpha\eta(\tau(p_\alpha))$;
\item{(E)} if $\fN_1,\ldots ,\fN_k$ are pairwise commuting
then $H(\fN_1,\ldots ,\fN_k)=H((\fN_1\cup\ldots \cup\fN_k)
^{\prime\prime})$;
\item{(F)} if $\Phi$ is an automorphism of $\fR$ preserving
the trace $\tau$, then $H(\Phi(\fN_1),\ldots ,\Phi(\fN_k))=
H(\fN_1,\ldots ,\fN_k)$.

Now, if $\Phi$ is a $\tau$-preserving automorphism of $\fR$, then
properties (B) and (F) imply that that the limit
$$
H(\fN,\Phi)=\lim_{k\to\infty}{1\over k}H(\fN,\Phi(\fN),\ldots ,
\Phi^{k-1}(\fN))\ref{\entropyref}
$$
exists. We define the CS entropy as the supremum
of the above quantity over all possible choices of the finite
dimensional algebra $\fN$,
$$
h_{CS}(\Phi)=\sup_{\fN,\;\dim\fN<\infty}H(\fN, \Phi).\ref{\costref}
$$
To be able to compute $h_{CS}(\Phi)$ we need a non-commutative version
of the Kolmogorov-Sinai theorem. Such a theorem was proved in [CS]
and is formulated as follows.

\thm\connstorthm{Let $\{\fN_k\}$ be an increasing sequence of
finite dimensional von Neumann subalgebras of $\fR$ such that
the weak closure $\big(\bigcup_k\fN_k\big)^-$ of $\bigcup_k\fN_k$
is $\fR$. Then
$$
h_{CS}(\Phi)=\lim_{k\to\infty}H(\fN_k,\Phi).\ref{\entlimitref}
$$
}

Recall that von Neumann algebras having the property assumed in
the theorem above are called hyperfinite. This theorem was used in
[CS] to compute the entropy of the non-commutative Bernoulli shift.

As expected, the CS entropy reduces to the KS entropy in the
commutative case.

\thm\abelthm{Let $M$ be a compact space with a Borel probability
measure $d\nu$ and let $\varphi$ be a measure preserving automorphism
of $M$. Consider the von Neumann algebra $\fR=L^\infty(M)$ with the
trace $\tau$ given by $\tauabeldef$, and the automorphism $\Phi$ of
$\fR$ defined by \Phidefref . Then $h_{CS}(\Phi)=h_{KS}(\varphi)$.}

\subsec
The actual definition of $H(\fN_1,\ldots ,\fN_k)$ will play a
role below and so we summarize it briefly.

We consider a von Neumann subalgebra $\fN\subset\fR$, and define
the following inner product on $\fN$: $(x,y)=\tau(x^*y)$.
The completion of $\fN$	in the norm induced by this inner product
is a Hilbert space which we denote by $L^2(\fN)$. Let $P_\fR:
L^2(\fR)\rightarrow L^2(\fN)$ be the orthogonal projection on
$L^2(\fN)$ and let $E_\fN$ denote the restriction of $P_\fN$
to the dense subspace $\fR\subset L^2(\fR)$. This is a
non-commutative version of the conditional expectation operator.

Let now $\cS_k$ be the set of all sequences of elements of $\fR$,
$x=\{x_{\bf i}\}$, where ${\bf i}\in\bN^k$, such that:
\item{(a)} $x_{\bf i}\geq 0$;
\item{(b)} all but finitely many $x_{\bf i}$ are zero;
\item{(c)} $\sum_{\bf i}x_{\bf i}=I$.

\noindent
For $x\in\cS_k$ and $1\leq l\leq k$ we set
$$
x^l_j=\cases{x_j,&if $k=1$;\cr
             \phantom{0}\cr
             \sum_{i_1\ldots i_{l-1}i_{l+1}\ldots i_k}
              x_{i_1\ldots i_{l-1}ji_{l+1}\ldots i_k},
               &if $k\geq 2$.\cr}\num
$$
We define
$$
H(\fN_1,\ldots ,\fN_k)=\sup_{x\in\cS_k}\big\{\sum_{{\bf i}\in
\bN^k}\eta(\tau(x_{\bf i}))-\sum_{l,j}\tau(\eta(E_{\fN_l}x^l_j))
\big\}.\ref{\entropydefref}
$$
It now takes quite a lot of skill to establish the results stated
above, and we refer the interested reader to [CS] for details.

\subsec
We now formulate and prove a technical result which will be a
basis for the arguments of next section.

\lemma\tensprodlemma{Let $\fR_1=L^\infty(M)$, where $M$ is a
compact space with a Borel probability measure $d\nu$ and the
natural faithful normal trace $\tau_1(\cdot)=\int_M(\cdot)d\nu$,
let $\fR_2$ be a finite dimensional von Neumann algebra with
a faithful normal trace $\tau_2$, and let $\Psi$ and $\Phi$ be trace
preserving automorphisms of $\fR_1$ and $\fR_2$, respectively.
Consider the hyperfinite von Neumann algebra $\fR=
\fR_1\otimes\fR_2$ with the faithful normal trace $\tau = \tau_1
\otimes\tau_2$, and the $\tau$-preserving automorphism $\Gamma
= \Psi\otimes\Phi$ of $\fR$. Then $h_{CS}(\Gamma)=h_{CS}(\Psi)$.}

\medskip\noindent
{\it Proof.} The proof of this lemma proceeds in steps.

\noindent{\it Step 1.} For any collection of finite dimensional
subalgebras $\fN_1,\ldots ,\fN_k\subset\fR_1$,
$$
H(\fN_1\otimes\fR_2,\ldots ,\fN_k\otimes\fR_2)=
H((\fN_1\cup\ldots\cup\fN_k)^{\prime\prime}\otimes\fR_2).
\ref{\steponeref}
$$
To prove this, note first that by property ($C$) of Section
\connessec ,
$$
H(\fN_1\otimes\fR_2,\ldots ,\fN_k\otimes\fR_2)\leq
H((\fN_1\cup\ldots\cup\fN_k)^{\prime\prime}\otimes\fR_2),\num
$$
as $\fN_j\otimes\fR_2\subset(\fN_1\cup\ldots\cup\fN_k)^{\prime\prime}
\otimes\fR_2$. To prove that
$$
H(\fN_1\otimes\fR_2,\ldots ,\fN_k\otimes\fR_2)\geq
H((\fN_1\cup\ldots\cup\fN_k)^{\prime\prime}\otimes\fR_2),\num
$$
we proceed as follows. Let $P^j_1,\ldots , P^j_{n_j}$, $1\leq
j\leq n_j$, where $n_j={\rm dim}\fR_j$, denote the minimal
projections in $\fN_j$, and let $E_1,\ldots , E_n$ be minimal
projections in $\fR_2$ such that $\sum_jE_j=I$.	We set
$$
x_{i_0i_1\ldots i_k}=E_{i_0}\otimes P_{i_1}\cdot\ldots\cdot
P_{i_k}=(E_{i_0}\otimes P_{i_1})\cdot\ldots\cdot(E_{i_0}\otimes
P_{i_k}),\ref{\xdefref}
$$
and observe that $\{x_{i_0i_1\ldots i_k}\}\in S_{k+1}$ and it
forms a system of minimal projections in $(\fN_1\cup\ldots\cup\fN_k)
^{\prime\prime}\otimes\fR_2$. Since $\eta(x_{i_0i_1\ldots i_k})=0$,
property ($D$) of Section \connessec\ implies that
$$
\eqalign{
H((\fN_1\cup\ldots\cup\fN_k)^{\prime\prime}\otimes\fR_2)
&=\sum_{i_0i_1\ldots i_k}\tau(\eta(x_{i_0i_1\ldots i_k}))\cr
&\leq H(\fN_1\otimes\fR_2,\ldots ,\fN_k\otimes\fR_2).\cr
}
$$

\noindent{\it Step 2.} If $\fN$ is a finite dimensional subalgebra
of $\fR_1$, then
$$
H(\fN\otimes\fR_2)=H(\fN)+H(\fR_2).\ref{\additref}
$$
To prove this, note that for a projection $P\in\fN_1$ and a
projection $E\in\fR_2$,
$$
\tau(\eta(P\otimes E))=\tau_1(\eta(P))\tau_2(E)+
\tau_1(P)\tau_2(\eta(E)).\ref{\factpropref}
$$
Denoting by $P_1,\ldots ,P_m$ and $E_1,\ldots ,E_n$ systems of
minimal projections in $\fN$ and $\fR_2$, respectively, and
using property ($D$), we obtain
$$
\eqalign{
H(\fN\otimes\fR_2)&=\sum_{j,k}\tau_1(\eta(P_j))\tau_2(E_k)+
\tau_1(P_j)\tau_2(\eta(E_k))\cr
&=\sum_j\tau_1(\eta(P_j))+\sum_k\tau_2(\eta(E_k))\cr
&=H(\fN)+H(\fR_2).\cr
}
$$

\noindent{\it Step 3.} Choose now an increasing sequence
$\{\fP_n\}_{n\in\bN}$ of finite dimensional subalgebras of
$\fR_1$, such that $\big(\bigcup_n\fP_n\big)^-=\fR_1$. Then
$\{\fP_n\otimes\fR_2\}_{n\in\bN}$ forms an increasing
sequence of finite dimensional subalgebras of $\fR_1\otimes
\fR_2$, and $\big(\bigcup_n\fP_n\otimes\fR_2\big)^-=\fR_1\otimes
\fR_2$.	Therefore, by \connstorthm ,
$$
h_{CS}(\Psi\otimes\Phi)=\lim_{n\to\infty}H(\fP_n\otimes\fR_2,
\Psi\otimes\Phi).\num
$$
By Steps 1 and 2,
$$
\eqalign{
H(\fP_n\otimes\fR_2,\; &\Psi(\fP_n)\otimes\Phi(\fR_2), \ldots ,
\Psi^{k-1}(\fP_n)\otimes\Phi^{k-1}(\fR_2))\cr
&=H(\fP_n\otimes\fR_2,\Psi(\fP_n)\otimes\fR_2, \ldots ,
\Psi^{k-1}(\fP_n)\otimes\fR_2)\cr
&=H((\fP_n\cup\Psi(\fP_n)\cup\ldots\cup\Psi^{k-1}(\fP_n))
^{\prime\prime}\otimes\fR_2)\cr
&=H((\fP_n\cup\Psi(\fP_n)\cup\ldots\cup\Psi^{k-1}(\fP_n))
^{\prime\prime})+H(\fR_2)\cr
&=H(\fP_n,\Psi(\fP_n),\ldots ,\Psi^{k-1}(\fP_n))+H(\fR_2).\cr
}
$$
But $H(\fR_2)$ is a constant independent of $k$, and so
$$
H(\fR_1\otimes\fR_2,\Psi\otimes\Phi)=H(\fR_1,\Psi),\num
$$
which proves the lemma.	$\square$

\section\entropysec{Entropy of the quantized dynamics}

\newsubsec
We are now ready to compute the CS entropy of the quantized cat
and baker's dynamics.

\thm\catentthm{The CS entropy of the quantized cat dynamics on
the torus is equal to the classical value,
$$
h_{CS}(\Gamma_\hbar)=\log|\mu_1|.\ref{\entropyvalref}
$$
Furthermore, if $|{\rm tr}(\gamma)|\leq 2$, then
$h_{CS}(\Gamma_\hbar)=0$.}

\medskip

It is an interesting question, even if without physical
significance, whether \catentthm\ holds without the assumption
that $\lambda/2\pi$ is rational. In that case, $\fA_\hbar$ is not
isomorphic to a finite dimensional algebra tensored by an abelian
algebra, and so \tensprodlemma\ cannot be applied. In the case of
topological entropy, Voiculescu [V1] has recently shown that the
entropy of the quantized dynamics does not exceed the classical
value.

An analogous result holds for the quantized baker's map.

\thm\bakentthm{The CS entropy of the quantized baker's map
is equal to the KS entropy of the classical baker's map,
$$
h_{CS}(\bak)=\log 2.\num
$$}

\medskip
It is easy to prove the above theorems. Indeed, according to
\tensprodthm\ and \autfactthm , $\fA_\hbar$ and $\Gamma_\hbar$
have precisely the structure required by \tensprodlemma .
Hence, $h_{CS}(\Gamma_\hbar)=h_{CS}(\Psi_\hbar)$. It is easy
to see that the map $\theta\rightarrow\gamma\theta+\Delta_\gamma$
is conjugate to the cat map $\theta\rightarrow\gamma\theta$.
According to the well known theorem [CFS], conjugate maps
have equal KS entropies, and so \abelthm\ implies
that $h_{CS}(\Phi)=h_{KS}(\gamma)$. \catentthm\ follows
from \catentref .

The proof of \bakentthm\ is analogous, with \factorthm\
replacing \autfactthm , and the final conclusion following
from \bakentref . $\square$

\subsec
We conclude this section with a brief discussion of a dynamical
system on a torus which is ergodic but is not chaotic. Consider
the Kronecker map on the torus defined by
$$
K:\; (x_1,x_2)\to (x_1+\omega_1,x_2+\omega_2).\num
$$
This map is known to be ergodic if and only if the frequencies
$\omega_1$ and $\omega_2$ are linearly independent over $\bZ$.
The Kronecker map is, however, not chaotic, as its KS entropy
is easily found to be zero [CFS].

In terms of the complex variable $z$, the Kronecker map reads
$$
K\; :z\to z+\omega,
$$
with $\omega =(\omega_1+i\omega_2)/\sqrt 2$, and so to quantize
it we need to find a unitary operator implementing the
following Bogolubov transformation:
$$
A^\dagger\to A^\dagger+\omega I\ .
$$
As in the case of the cat map the unitary operator is
uniquely (up to a phase) determined by the above condition.
In fact, an easy consequence of \udefref\ is that
$$
U(-\omega)A^\dagger U(-\omega)^{-1}=A^\dagger +\omega I,
$$
and so $U(-\omega)$ is the required unitary operator.

Let now $K_\hbar$ be the automorphism of the quantum torus given by
by $K_\hbar(\cdot)=U(-\omega)(\cdot)U(-\omega)^{-1}$. Evaluated on
the generators of $\fA_\hbar$, $K_\hbar$ is:
$$
\eqalign{
K_\hbar(U)&=e^{2i\pi\omega_1}U,\cr
K_\hbar(V)&=e^{2i\pi\omega_2}V.\cr
}\ref{\kronref}
$$

Assume now that $\hbar=1/ 2\pi N$, in which case \tensprodthm\ is
applicable. It is easy to see that $K_\hbar$ can be factorized,
with the first factor given by the following automorphism of
$L^\infty(\bT^2)$:
$$
f(\theta)\to f(\theta_1 +N\omega_1, \theta_2 +N\omega_2).
\ref{\firstfactref}
$$
Hence, the CS entropy of $K_\hbar$ is equal to the KS entropy of
\firstfactref\ and is thus zero.

\medskip\noindent
{\bf Acknowledgment.} We would like to thank Neepa Maitra and
Ron Rubin for very helpful remarks. We would also like to thank
an anonymous referee of this paper for his constructive and very
helpful comments.

\vfill\eject

\centerline{\bf References}
\baselineskip=12pt
\frenchspacing

\bigskip

\item{[A]} Arnold, V.: {\it Geometrical Methods in the Theory of
Ordinary Differential Equations}, Springer Verlag (1983)
\item{[AW]} Adler, R. L., and Weiss, B.: Similarity of automorphisms
of the torus, {\it Memoirs AMS}, {\bf 98} (1970)
\item{[BV]} Balazs, N. L., and Voros, A.: The quantized baker's
map, {\it Ann. Phys.}, {\bf 190}, 1--31 (1989)
\item{[B1]}Berezin, F. A.: General concept of quantization, {\it Comm.
Math. Phys.}, {\bf 40}, 153--174 (1975)
\item{[B2]} Berry, M. V.: Quantum chaology, {\it Proc R. Soc. Lond.},
{\bf A413}, 183-198 (1987)
\item{[B3]} Brown, J. R.: {\it Ergodic Theory and Topological
Dynamics}, Academic Press (1976)
\item{[BC1]} Berger, C. A., and Coburn, L. A.: Toeplitz operators
and quantum mechanics, {\it J. Funct. Anal.}, {\bf 68}, 273--299
(1986)
\item{[BC1]} Berger, C. A., and Coburn, L. A.: Toeplitz operators
on the Segal - Bargmann space, {\it Trans AMS}, {\bf 301}, 813--829
(1987)
\item{[BD]} Bouzouina, A., and De Bievre, S.: Equipartition of
the eigenfunctions of quantized ergodic transformations of the
torus, {\it Comm. Math. Phys.}, to appear
\item{[BDG]} de Bievre, S., Degli Esposti, M., and Giachetti, R.:
Quantization of a piecewise affine transformations on the torus,
preprint (1994)
\item{[CS]} Connes, A., and St\o rmer, E.: Entropy for automorphisms
of $II_1$ von Neumann algebras, {\it Acta Math.}, {\bf 134}, 289--306
(1975)
\item{[CNT]} Connes, A., Narnhofer, H., and Thirring, W.: Dynamical
entropy of $\bC^*$-algebras and von Neumann algebras, {\it Comm.
Math. Phys.}, {\bf 112}, 691--719 (1987)
\item{[CFS]} Cornfeld, I. P., Fomin, S. V., and Sinai, Ya.:
{\it Ergodic Theory}, Springer Verlag (1982)
\item{[DGI]} Degli Esposti, M., Graffi, S., and Isola, S.:
Classical limit of the quantized hyperbolic toral automorphisms,
{\it Comm. Math. Phys.}, {\bf 167}, 471--507 (1995)
\item{[D1]} Degli Esposti, M.: Quantization of the orientation
preserving automorphisms of the torus, {\it Ann. Inst. H. Poincare},
{\bf 58}, 323--341 (1993)
\item{[D2]} Dixmier, J.: {\it Von Neumann Algebras}, North Holland
(1981)
\item{[F]} Folland, G.: {\it Harmonic Analysis in Phase Space},
Princeton University Press (1988)
\item{[HB]} Hannay, J. J., and Berry, M. V.: Quantization of linear
maps on a torus - Fresnel diffraction by a periodic grating,
{\it Physica} {\bf D1},	267--291 (1980)
\item{[HT]} Heller, E. J. and Tomsovic, S.: Postmodern quantum
mechanics, {\it Phys. Today}, 38--46 (1993)
\item{[K1]} Keating, J. P.: Asymptotic properties of the periodic
orbits of the cat map, {\it Nonlinearity}, {\bf 4}, 277--307 (1990)
\item{[K2]} Keating, J. P.: The cat maps: quantum mechanics and
classical motion, {\it Nonlinearity}, {\bf 4}, 309--341 (1990)
\item{[KL]} Klimek, S., and Le\'sniewski, A.: Quantum Riemann surfaces,
III. The exceptional cases, {\it Lett. Math. Phys.}, {\bf 32}, 45--61
(1994)
\item{[MO]} de Matos, M. B., and Ozorio de Almeida, A. M.:
Quantization of Anosov maps, {\it Ann. Phys.}, {\bf 237}, 46--65 (1995)
\item{[N]} Nakamura, K.: {\it Quantum Chaos. A New Paradigm of
Nonlinear Dynamics}, Cambridge University Press (1993)
\item{[OTH]} O'Connor, P. W., Tomsovic, S., and Heller, E. J.:
Accuracy of semiclassical dynamics in the presence of chaos,
{\it J. Stat. Phys.}, {\bf 68}, 131--152 (1992)
\item{[R1]} Rieffel, M.: Deformation quantization of Heisenberg
manifolds, {\it Comm. Math. Phys.}, {\bf 122}, 531--562 (1989)
\item{[R2]} Rieffel, M.: Non-commutative tori - a case study of
non-commutative differentiable manifolds, {\it Contemp. Math.},
{\bf 105}, 191--211 (1990)
\item{[S]} Saraceno, M.: Classical structures in the quantized
baker transformation, {\it Ann. Phys.}, {\bf 199}, 37--60 (1990)
\item{[SV]} Saraceno, M., and Voros, A.: Towards a semiclassical
theory of the quantum baker's map, {\it Physica}, {\bf D 79},
206--268 (1994)
\item{[V1]} Voiculescu, D.: Dynamical approximation entropies
and topological entropy in operator algebras, {\it Comm. Math.
Phys.}, to appear
\item{[V2]} Voros, A.: Aspects of semiclassical theory in the
presence of classical chaos, {\it Prog. Theor. Phys., Suppl.}
{\bf 116}, 17--44 (1994)
\item{[Z]} Zelditch, S.: Quantum ergodicity of quantized contact
transformations and ergodic symplectic toral automorphisms,
preprint (1995)

\vfill\eject\end